\documentclass[twocolumn,aps,amsmath, amssymb,longbibliography]{revtex4-1}
\usepackage{graphicx}
\usepackage{amsmath}
\usepackage{amssymb}
\usepackage{color}
\usepackage{natbib}
\usepackage[pdftex,breaklinks,bookmarks=false,colorlinks,linkcolor=blue,citecolor=blue,urlcolor=blue]{hyperref}

\begin{document}
\title{Quantized Nonlinear Conductance in Ballistic Metals}

\author{C. L. Kane}
\affiliation{Department of Physics and Astronomy, University of Pennsylvania, Philadelphia, PA 19104}

\begin{abstract}
We introduce a non-linear frequency dependent $D+1$ terminal conductance that characterizes a $D$ dimensional Fermi gas, generalizing the Landauer conductance in $D=1$.  For a 2D ballistic conductor we show that this conductance is quantized and probes the Euler characteristic of the Fermi sea.    
We critically address the roles of electrical contacts and of Fermi liquid interactions, and we propose experiments on 2D Dirac materials such as graphene using a triple point contact geometry. 
\end{abstract}

\maketitle

A dramatic consequence of the role of topology in the structure of quantum matter is the existence of topological invariants that are reflected in quantized response functions.   The hallmark of this is the integer quantized Hall effect (IQHE)\cite{vonklitzing80}, which probes the Chern number characterizing the topology of a gapped 2 dimensional (2D) electronic phase\cite{tknn82}.   Quantum topology also plays a role in the electrical response of metals.   For example, the Berry phase associated with the Fermi surface of a 2D metal contributes to an intrinsic non-quantized part of the anomalous Hall conductivity\cite{haldane04}.   In 3D, the Chern number associated with the Fermi surface in a Weyl semimetal leads to a quantized circular photogalvanic effect\cite{dejuan17} in the absence of disorder and interactions\cite{avdoshkin20}.   In addition to the quantum topology associated with the twisting of the quantum states on the Fermi surface, metals also exhibit a simpler geometric topology associated with the Fermi surface.   It is well known that noble metals, like copper, have a Fermi surface with a non-trivial genus\cite{ashcroft76}.  While Fermi surfaces have been mapped in detail, and Lifshitz transitions\cite{lifshitz60} where their topology changes have been characterized\cite{volovik17,volovik18}, Fermi surface topology has not been measured directly.  Here we pose the question of whether the topology of the Fermi surface is associated with a quantized response.

An indication that the answer 
is affirmative is provided by the 1D case.  The Landauer conductance of a ballistic 1D conductor is $e^2/h$ times the number of occupied bands\cite{landauer57,fisher81}.   While this quantization is related to the IQHE, 
there are important differences.   First it is less robust, since it relies on reflectionless contacts and the absence of scattering.   Nonetheless, conductance quantization has been observed in quantum point contacts\cite{vanwees88}, 1D semiconductor wires\cite{honda95,vanweperen13} and carbon nanotubes\cite{frank98}, albeit with less precision than the IQHE.  A second difference is that unlike the IQHE, the quantized value does not reflect the topology of a 2D gapped state, but rather the topology of the 1D filled Fermi sea.    

In this paper, we seek to generalize this to higher dimensions.  For a $D$ dimensional ballistic conductor with suitably defined ideal leads, we introduce a $D+1$ terminal frequency dependent nonlinear conductance,
\begin{equation}
I_{D+1}(\omega_\Sigma) = G(\{\omega_p\}) \prod_{p=1}^D V_p(\omega_p),
\label{nonlinear}
\end{equation}
where $I_p(\omega_p)$ ($V_p(\omega_p)$) are the current (voltage) in lead $p$ at frequency $\omega_p$ and $\omega_\Sigma = \sum_p \omega_p$.   We will show that for $D=1$ and $2$, $G(\{\omega_p\})$ has a universal term of the form,
\begin{equation}
G(\{\omega_p\}) = \frac{i\omega_\Sigma}{\prod_p (i \omega_p)} \frac{e^{D+1}}{h^D} \chi_F,
\label{g}
\end{equation}
where $\chi_F$ is the {\it Euler characteristic} of the $D$-dimensional Fermi sea.  We will focus on $D=2$, leaving the generalization to $D>2$ to future work.   This result will be established for non-interacting electrons by first presenting a simple thought experiment, which is formalized by a semi-classical Boltzmann transport theory.   This will be followed by a more general quantum non-linear response theory, which reproduces the Boltzmann theory.   We will then critically assess the prospects for experimentally measuring $\chi_F$ in a 2D conductor using a {\it triple point contact}.   Crucial issues to be addressed include the role of electrical contacts and electron-electron interactions, which place bounds on the applicability of (\ref{g}).

The Euler characteristic is defined as\cite{nakahara90}
\begin{equation}
\chi_F = \sum_{l=0}^D (-1)^l b_l,
\label{bettief}
\end{equation}
where $b_l$ is the $l$'th Betti number, 
given by the rank of the $l$'th homology group, which counts the topologically distinct 
$l$-cycles.  In 1D, $\chi_F$ is the number of disconnected components of the Fermi sea.   In general, $\chi_F$ can be expressed as a sum over the disconnected components of the Fermi {\it surface}.
In 2D, electron-like, hole-like and open Fermi surfaces contribute  $+1$, $-1$ and $0$, respectively.  In 3D, each Fermi surface with genus $g_k$ contributes $1-g_k$.  Note that completely empty bands and completely filled bands both have $\chi_F=0$, and electron-like and hole-like Fermi surfaces have opposite sign for even $D$.

Morse theory\cite{milnor69,nash88} provides a representation of $\chi_F$ in terms of the critical points of the electronic dispersion $E({\bf k})$, where ${\bf v}_{\bf k} = \nabla_{\bf k} E({\bf k})/\hbar = 0$ for $E({\bf k}) < E_F$:
\begin{equation}
\chi_F = \sum_m \eta_m.
\label{morseef}
\end{equation}
$m$ labels the critical points (assumed non-degenerate) with signature $\eta_m = {\rm sgn}({\rm det}[\partial^2 E({\bf k}_m)/\partial k_i\partial k_j])$.   This shows that $\chi_F$ changes at a Lifshitz transition\cite{lifshitz60}, when a minimum, maximum or saddle point passes through $E_F$, signaling a change in Fermi surface topology.

To motivate our result, we review and then generalize a thought experiment\cite{laughlin81} that explains the quantization of the 1D Landauer conductance.   Consider an infinitely long 1D electron gas (1DEG), with electronic states $E(k)=\hbar^2 k^2/2m$ filled to $E_F$.   Apply a $h/e$ voltage pulse $V(t)$ by introducing a slowly varying electric field $E(x,t)$ that is nonzero near $x=0$ and $t=0$, such that $\int dx dt E(x,t) = h/e$.  This will lead to a charge $Q = G h/e$ transferred into the right lead, where $G$ is the conductance.   The charge $Q = e$ can be deduced from the fact that the impulse $V(t)$ transfers precisely one electron between the left- and right-moving Fermi points, reflecting the chiral anomaly associated with 1D chiral fermions.   The chiral anomaly is a consequence of the fact that the right and left movers are connected at the critical point $k=0$.  Due to the impulse one electron crosses $k=0$ and changes direction.   This argument can be generalized to a more complicated dispersion $E(k)$.   An electron will change direction at every critical point inside the Fermi sea, leading to a net transferred charge $Q = e \chi_F$, with $\chi_F$ given in (\ref{morseef}).   It follows that $G = \chi_F e^2/h$.

We now seek a version of this argument for $D=2$.  Consider a 2DEG with dispersion $E({\bf k})$ defined on an infinite plane that is divided into 3 regions that meet at a point.   Apply a $h/e$  pulse $V_1(t)$ to region $1$, followed by a $h/e$  
pulse $V_2(t)$ to region $2$ by introducing electric fields near their boundaries.  Each pulse will lead to a charge transferred to lead 3 that scales with the length of the contact.   However, we will argue that the excess charge $Q_3$, defined as the charge transferred due to the two pulses with the charge transferred for independent pulses subtracted off, will be universal and given by $Q_3 = e \chi_F$, where $\chi_F$ is the Euler characteristic of the 2D Fermi sea.   This excess charge defines a second order non-linear response that can be isolated in the frequency domain.

\begin{figure}
\includegraphics[width=3 in]{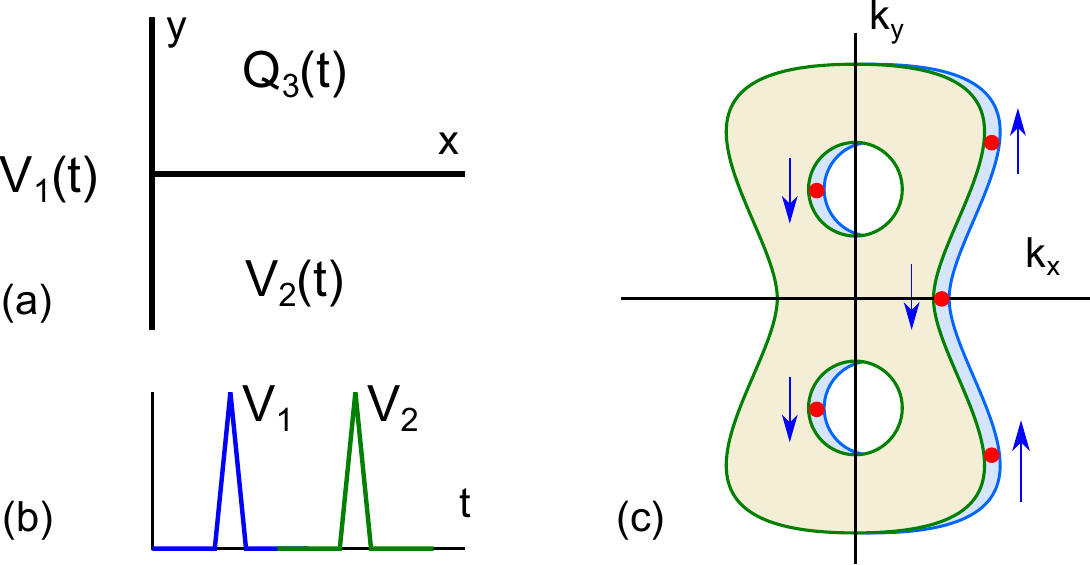}
\caption{(a,b) A thought experiment in which $h/e$ voltage pulses are applied to regions 1 and 2 in (a).  (c) shows a hypothetical 2D Fermi sea with $\chi_F = -1$.
After pulse $V_1(t)$ for $x>0$ there are extra electrons (shown in blue) propagating to the right.   Pulse $V_2(t)$ accelerates those electrons, and $v_y$ changes sign at the points indicated by the red dots in a direction indicated by the arrows. 
 The net excess charge in region 3 is determined by the difference between the number of concave and convex critical points on the Fermi surface with ${\bf v}_{\bf k} \propto +\hat x$, which measures $\chi_F$.}
\label{Fig1}
\end{figure}

It is simplest to consider the geometry in Fig. \ref{Fig1}a, where region 1 is the half-plane $x<0$, and regions 2 (3) are the quadrants $x>0$, $y<0$($y>0$).   In that case, the $V_1(t)$ pulse accelerates electrons in the $x$ direction, so that for every value of $k_y$ on the Fermi surface there is one extra electron propagating to the right ($v_x >0$) for $x>0$, and one extra hole propagating to the left ($v_x <0$) for $x<0$.   $Q_3$ will be determined by the effect of the $V_2(t)$ pulse on those extra electrons with $x>0$, which are accelerated in the $y$ direction.   As in the 1D example discussed above, the transferred charge can be determined by counting those electrons that change directions at the critical points on the Fermi surface where $v_x>0$ and $v_y=0$.   Referring to the hypothetical Fermi surface in Fig. 1c, these arise at the points indicated by the  dots, which come in two varieties distinguished by whether the Fermi surface is concave (convex) with $\partial^2 E/\partial k_y^2 >0$($<0$).  $Q_3$ is then $e$ times the sum over those critical points with signs ${\rm sgn}[\partial^2 E/\partial k_y^2]$.  By inspecting Fig. 1c, it is clear that this is $\chi_F$.  
This will be proven below.  We conclude that $Q_3 = e \chi_F$.

The above argument can be sharpened by developing a semi-classical Boltzmann transport theory.   In the absence of scattering the electron distribution function $f({\bf k},{\bf r},t)$ satisfies the collisionless Boltzmann equation,
\begin{equation}
(\partial/\partial t + {\bf v}_{\bf k}\cdot \nabla_{\bf r} + e {\bf E} \cdot \nabla_{\bf k}/\hbar)f = 0.
\label{boltzmann}
\end{equation}
Consider two weak pulses, ${\bf E}_1(x_1,y_1,t) = \xi_1 \delta(t-t_1) \delta(x_1) \hat x $ and 
${\bf E}_2(x_2,y_2,t) = \xi_2 \delta(t-t_2)[ \delta(y_2)\theta(x_2) \hat y - \delta(x_2) \theta(-y_2) \hat x]$.
We compute the charge
\begin{equation}
Q_3(t_3) =  e \int \frac{d^2 k}{(2\pi)^2} \int_0^\infty dx_3 dy_3 \delta f({\bf k},x_3,y_3,t_3)
\label{q3boltzmann}
\end{equation}
perturbatively at order $\xi_1 \xi_2$ for $t_1 < t_2 < t_3$.   Integrating (\ref{boltzmann}) to this order gives $\delta f = e^2 \xi_1 \xi_2 \delta\tilde f/\hbar^2$ with
\begin{equation}
\delta\tilde f
=   \int dy_1 dx_2
\delta({\bf r}_{32}-{\bf v}_{\bf k}t_{32}) \frac{\partial}{\partial{k_y}}[ \delta({\bf r}_{21}-{\bf v}_{{\bf k}} t_{21})  \frac{\partial f_0}{\partial{k_x}}],
\label{f2boltzmann}
\end{equation}
where $f_0({\bf k}) = \theta(E_F - E({\bf k}))$, $t_{ij}=t_i-t_j$, ${\bf r}_{32} = (x_3-x_2,y_3)$, ${\bf r}_{21} = (x_2,-y_1)$, and $x_2$ ($y_1$) are integrated from $0$ ($-\infty$) to $\infty$.  The second term in ${\bf E}_2$, with $x_2=0$, is absent because the $\delta$ functions can not be satisfied.   After plugging (\ref{f2boltzmann}) into (\ref{q3boltzmann}), the four spatial integrals cancel the $\delta$ functions, but since $x_2>0$ ($y_3>0$), we require $v_x>0$ ($v_y>0$) inside (outside) $\partial/\partial k_y$.  After integrating by parts on $k_y$ and replacing $\xi_{1,2}\rightarrow h/e$ we obtain
\begin{equation}
Q_3 = - e \int d^2k
\frac{\partial f_0({\bf k})}{\partial k_x} \theta(v_x({\bf k})) \frac{\partial \theta(v_y({\bf k}))}{\partial k_y} .
\label{q3fs}
\end{equation}
This captures the result of the heuristic argument above: $-\partial f_0/\partial k_x$ isolates the Fermi surface, while $\theta(v_x) \partial(\theta(v_y))/\partial k_y$ isolates the critical points on the Fermi surface identified in Fig. \ref{Fig1}c.    To make contact with Eq. \ref{morseef}, it is convenient to add zero to the integrand in the form
$-(\partial f_0/\partial k_y) \theta(v_x)\partial \theta(v_y)/\partial k_x $.   This is clearly zero, since $\partial f_0/\partial k_y = v_y \partial f_0/\partial E$, and $\partial\theta(v_y)/\partial k_x$ fixes $v_y=0$.   This allows us to integrate by parts to obtain,
\begin{equation}
Q_3 =  e \int d^2k f_0({\bf k}) [
\frac{\partial \theta(v_x)}{\partial k_x} \frac{\partial \theta(v_y)}{\partial k_y} 
-\frac{\partial \theta(v_x)}{\partial k_y} \frac{\partial \theta(v_y)}{\partial k_x} ]
\end{equation}
The integrand is only non-zero near critical points where $v_x=v_y=0$.  The integral evaluates the signature of each critical point, leading to
\begin{equation}
Q_3 = e \sum_m \eta_m = e \chi_F.
\label{q3chiF}
\end{equation}

We next consider the frequency domain response.   This can be computed using the Boltzmann theory, however, we will first formulate a more general quantum non-linear response theory, and show the Boltzmann theory follows, provided the applied fields vary slowly in space and time.
To this end, we introduce the Hamiltonian
\begin{equation}
{\cal H} = {\cal H}_0 + (V_1 \hat Q_1 e^{(\eta-i\omega_1)t} + V_2 \hat Q_2 e^{(\eta - i\omega_2)t} + {\rm h.c.})
\label{hq1q2}
\end{equation}
with ${\cal H}_0 = \sum_{\bf k} E({\bf k}) c_{\bf k}^\dagger c_{\bf k}$ and $\hat Q_p = \int d{\bf r} Q_p({\bf r}) \rho({\bf r})$ is defined in terms of the density operator $\rho({\bf r})$ for each of the three regions in Fig. 1a.   $Q_p({\bf r})$ is $1$ ($0$) inside (outside) region $p$ and is assumed to transition smoothly between $1$ and $0$ in a width $b$ near the boundary, with $k_F b \gg 1$.   

\begin{figure}
\includegraphics[width=3 in]{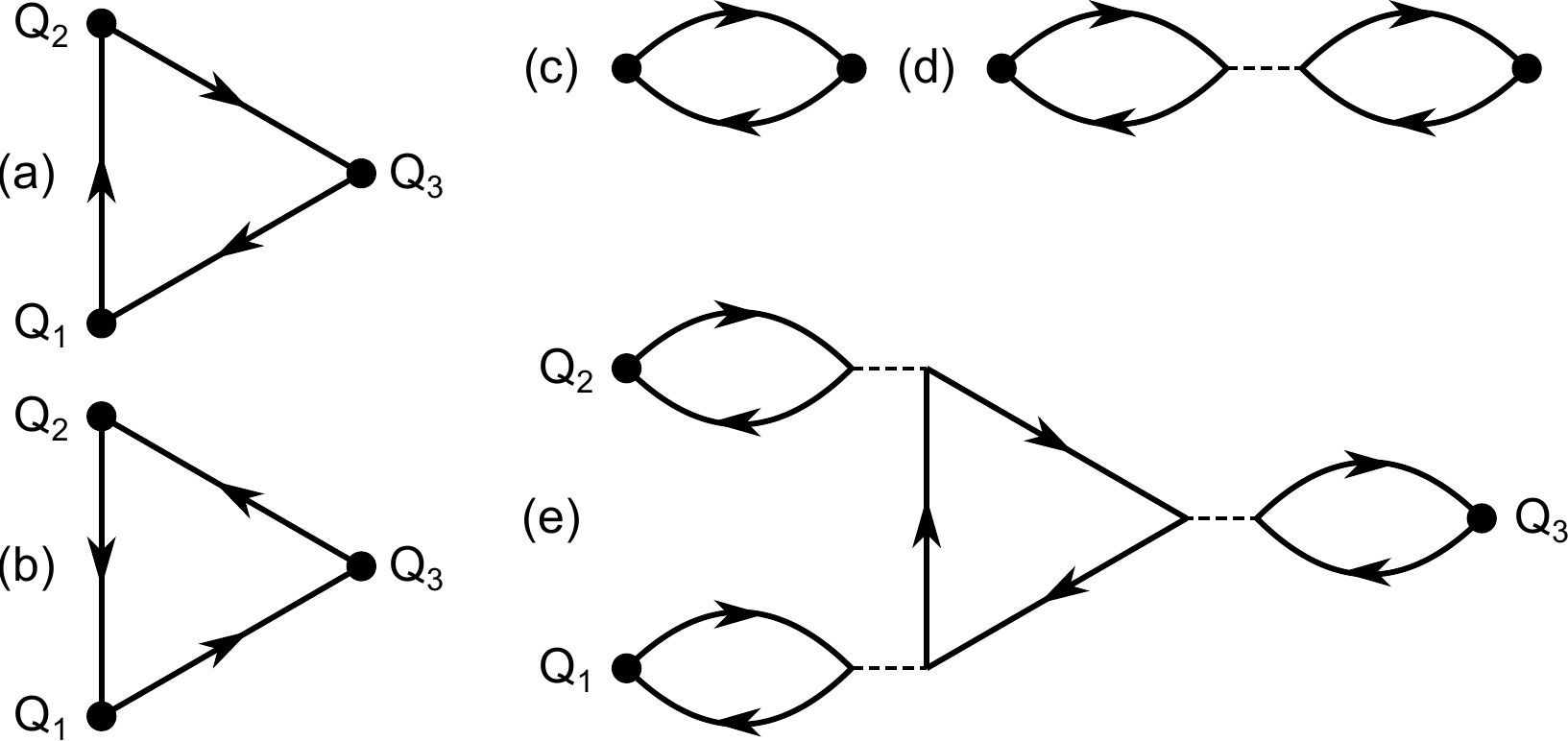}
\caption{Feynman diagrams for the conductance.  (a,b) describe the 2D non-linear conductance and (c) describes the 1D linear conductance in the absence of interactions and give the quantized conductance determined by $\chi_F$.  (d) and (e) show corrections to (a,b) and (c) that describe screening due to electron interactions and modify the quantized result.}
\label{Fig2}
\end{figure}

We compute the charge $Q_3(t)$ at frequency $\omega_1+\omega_2$ to order $V_1 V_2$.   We adopt a scalar potential formulation, which avoids the diamagnetic term.  The response has structure similar to  second order nonlinear optical response\cite{kraut79,vonbaltz81,zhang18}, and is determined by evaluating the Feynman diagrams in Fig. \ref{Fig2}a,b.
\begin{equation}
Q_3(\omega_1+\omega_2) = \alpha(\omega_1,\omega_2) V_1 V_2
\end{equation}
with
\begin{align}
\alpha(\omega_1,\omega_2) = &\frac{e^3}{\hbar^2} \sum_{l,m,n} \frac{f_l-f_m}{\omega_1-\omega_{lm}+i\eta}[
\frac{Q_1^{lm} Q_2^{mn} Q_3^{nl}}{ \omega_1+\omega_2-\omega_{ln}+i\eta}   \nonumber \\
&- \frac{Q_1^{lm} Q_3^{mn} Q_2^{nl}}{ \omega_1+\omega_2-\omega_{nm}+i\eta}  ] + (1\leftrightarrow 2).
\label{alphaw1w2}
\end{align}
Here $l,m,n$ label momenta, $\omega_{lm} = (E({\bf k}_l)-E({\bf k}_m))/\hbar$, $f_l = f_0({\bf k}_l)$ is a Fermi function, and 
\begin{equation}
Q_p^{lm} = \langle {\bf k}_l|\hat Q_p|{\bf k}_m\rangle = \int d{\bf r}_p Q_p({\bf r}_p) e^{i{\bf q}_{lm}\cdot{\bf r}_p}
\label{qalm}
\end{equation}
with ${\bf q}_{lm} = {\bf k}_l - {\bf k}_m$.  
Since $Q_p({\bf r})$ varies on the scale of $b$, an expansion in ${\bf q}_{lm}$ and ${\bf q}_{nm}$ is justified.  As shown in supplemental section \ref{Appendix A}, the ${\bf q}_{lm}$ and ${\bf q}_{nm}$ integrals can be performed to obtain,
\begin{align}
\alpha = &\int \frac{d^2 {\bf k} d^6{\bf r}_{1,2,3}}{(2\pi)^2}  \nabla^a_{\bf r} Q_1({\bf r}_1) \nabla^b_{\bf r} Q_2({\bf r}_2) Q_3({\bf r}_3)  [\nabla_{\bf k}^a f_0({\bf k})  \nonumber\\
& D({\bf r}_{21}, {\bf k},\omega_1) \nabla_{\bf k}^b D({\bf r}_{32},{\bf k},\omega_1+\omega_2)] 
+ (1\leftrightarrow 2)
\label{finalalpha}
\end{align}
with
$D({\bf r},{\bf k},\omega) = e^{-(\eta - i\omega)|{\bf r}|/|{\bf v}_{\bf k}|} \delta({\bf r} \times {\bf v}_{\bf k}) \theta({\bf r}\cdot {\bf v}_{\bf k})$.   
This form of the response also follows from solving (\ref{boltzmann}) in the frequency domain with ${\bf E}_p({\bf r},t) = -\nabla Q_p({\bf r}) V_p e^{(\eta-i\omega_p)t}$.

\begin{figure}
\includegraphics[width=2 in]{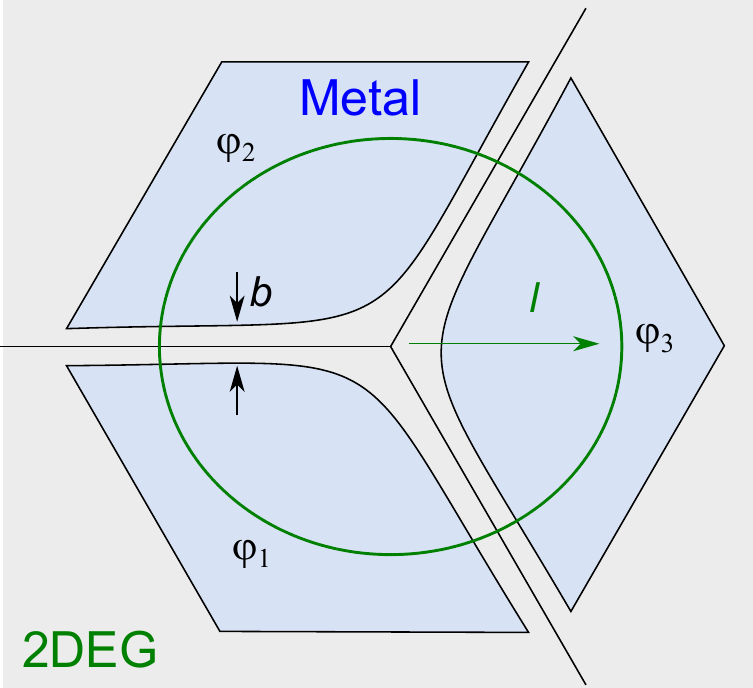}
\caption{A triple point contact as a model experimental geometry.   A 2D electron gas is connected to three metallic leads of size larger than the tunneling mean free path $\ell$.
The leads are separated by $b$ and subtend angles $\varphi_{1,2,3}$.  }
\label{Fig3}
\end{figure}

In supplemental section \ref{Appendix B}, we evaluate (\ref{finalalpha}) for the infinite plane in which three rays separate regions that subtend angles $\varphi_p$ (see Fig. \ref{Fig3}).  We show that there is an intrinsic term $\alpha_i(\omega_1,\omega_2) = \chi_F(e^3/h^2) /(\eta-i\omega_1)(\eta-i\omega_2)$ that is independent $\varphi_p$ (provided all $\varphi_p < \pi$\footnote{If one of the angles $\varphi_p$ is greater than $\pi$, $\alpha_i$ is still quantized, but its value is modified.   See supplemental section \ref{Appendix B}, where it is also established that for Fig. 1a the pulse argument analysis remains valid. }) as well as the detailed spatial profile of the fields.    In addition, there is an extrinsic term with a distinct frequency dependence, $\alpha_e(\omega_1,\omega_2) = k (e^3/h^2)/(\eta-i(\omega_1+\omega_2))^2$ with a coefficient $k$ that depends on $\varphi_p$ as well as the details of the Fermi surface.   The extrinsic term was not picked up by the pulse argument (where we assumed $t_2>t_1$), since it arises when ${\bf E}_1$ and ${\bf E}_2$ coincide in time.    The intrinsic term dominates when $\omega_1+\omega_2 \gg \omega_1$ or $\omega_2$.   The intrinsic term, which is exact for the infinite plane and non-interacting electrons is the central result of this paper.  To address experimental feasibility
we must consider the role of contacts as well as electron-electron interactions.
These both introduce complications into the analysis. 

As a model for electrical contacts, consider Fig. \ref{Fig3}, which depicts a 2D electron gas (2DEG) of size $L$ with large area tunnel contacts to ideal metallic leads separated by a distance $b$.  Provided the capacitance between the contacts and the 2DEG is sufficiently large, the voltages in the leads will establish a potential profile in the 2DEG as in Eq. \ref{hq1q2}.   We assume the tunnel barrier is in a regime in which the mean free path $\ell$ for tunneling from the 2DEG to the leads satisfies $k_F^{-1} \ll \ell \ll L$.  This defines a dwell time $\tau = \ell/v_F$ for electrons in the 2DEG.  In the pulse construction, we clearly require $t_{21}<\tau$, since for $t_{21}>\tau$ the first pulse has disappeared before the second arrives.   In the frequency domain calculation, tunneling to the leads introduces an exponential decay $e^{-|{\bf r}|/\ell}$ into $D({\bf r},{\bf k},\omega)$, which cuts off the small $\omega_1$ divergence in $\alpha$, effectively replacing $\omega_1 \rightarrow \omega_1+i/\tau$.  The coupling to the leads therefore places a lower bound $\omega_1,\omega_2\gg \omega_c$ for the applicability of (\ref{g}) with $\omega_c \sim \tau^{-1}$.  The intrinsic behavior is thus recovered when $\omega_c\ll \omega_{1(2)} \ll \omega_{2(1)}$.

A second complication involves the role of electron-electron interactions.  In 1D, the analog of our calculation is the Kubo formula conductance $G_{\rm Kubo}$, which is given by $G_{\rm Kubo} = K e^2/h$, where $K<1$ is the Luttinger parameter characterizing repulsive electron interactions\cite{apel82,kane92}.  However, $G_{\rm Kubo}$ does not correctly account for the electrical contacts, and it was argued that for Fermi liquid leads on a 1DEG of length $L$, $G(\omega) =e^2/h$ for $\omega\ll v_F/L$\cite{maslov95,ponomarenko95,safi95,safi97,safi99,thomale11}.  An appealing interpretation of this was explained by Kawabata\cite{kawabata95}, who argued that $G_{\rm Kubo}$ computed in an infinite system is renormalized because it describes the response to the {\it applied} voltage, while the quantized conductance, which reflects the chiral anomaly, is the response to the {\it self-consistent} potential, which includes screening due to  interactions.   Moreover, for the DC conductance it is the self-consistent potential that determines the measured electrochemical potential difference.  Shimizu\cite{shimizu96} emphasized the similarity of this description to Fermi liquid theory\cite{pines66}.   In terms of Feynman diagrams, for the response to the applied field, the Kubo formula bubble diagram (Fig. \ref{Fig2}(c)) is dressed by convolving with RPA-like polarization bubbles (Fig. \ref{Fig2}(d)), $\Pi(x,\omega)$, represented here in real space.   For a 1DEG in which leads are modeled by setting the interactions to zero for $|x|>L$, it can be checked that since $\Pi(x,\omega) \sim (\omega/v_F)e^{i \omega x/v_F}$, and $|x| < L$ the interaction corrections vanish for $\omega \ll v_F/L$.

To incorporate electron interactions into the calculation of the 2D non-linear response we adopt a renormalized Fermi liquid description\cite{shankar94} in which quasiparticles near $E_F$ interact with an energy $f_{{\bf k}{\bf k}'} n_{\bf k} n_{{\bf k}'}$, where 
$n_{\bf k} = c_{\bf k}^\dagger c_{\bf k}$.  At low energy, the important interaction corrections involve RPA bubbles\cite{rostami17}, and summing the diagrams like \ref{Fig2}(e) is equivalent to incorporating the Fermi liquid interactions $f_{{\bf k}{\bf k}'}$ into the Boltzmann equation\cite{pines66,shankar94}.   This is difficult to solve in general, but by evaluating the simplest diagram at first order in $f_{{\bf k}{\bf k'}}$ it can be checked that the response to the applied field is modified:  
$\chi_F \rightarrow \chi_F + O( N(E_F) f)$, where $f$ is an average of $f_{{\bf k}{\bf k}'}$ over the Fermi surface.  Thus, the non-linear response in 2D is modified by the Fermi liquid parameters just as the linear response in 1D is modified by the Luttinger parameter.   The bubble $\Pi(\omega,{\bf r}) \propto (\omega/v_F) e^{i\omega |{\bf r}|/v_F}$ vanishes for $\omega\ll \omega_c$ as for $D=1$.   However, since we must consider $\omega_1, \omega_2 \gg \omega_c$, the interaction corrections remain.   The origin of the correction is the same as in $D=1$: due to interactions, the potential is screened, as is accounted for by the RPA bubbles.   

At finite frequency, more detailed modeling is required to determine the relation between the self-consistent potential and the measured voltage.   In the absence of that, it will be fruitful to consider weakly interacting systems.   Consider a 2D Dirac material, such as graphene, with density of states $N(E) \sim |E|/v_F^2$.  For a short ranged interaction (screened by the leads) $N(E_F) f \ll 1$ for sufficiently small $E_F$.   Interestingly, the Fermi surface is electron-like (hole-like) for $E_F>0$($E_F<0$), so the response characterized by $\chi_F = 4 {\rm sgn} E_F$  (including spin and valley) changes sign at charge neutrality.

Our analysis opens several avenues for further inquiry.   On the practical side, it will be interesting to search for other measurable quantities that probe $\chi_F$.   Promising candidates include low frequency current noise as well as thermal conductance.   
It will also be interesting to generalize our theory to $D>2$ and to explore ways in which $\chi_F$ provides a fundamental characterization of a degenerate Fermi gas.   Our analysis suggests that $\chi_F$ defines a kind of ``higher order" anomaly.  In 1D, the chiral anomaly characterizes the connection between left and right moving electrons, which leads to a  lack of conservation of the right movers.  For $D>1$, $\chi_F$ characterizes a more general violation of conservation due to the fact that electrons propagating in different directions are connected at critical points ${\bf k}_m$.  
This is related to the Fermi surface anomaly discussed in Ref. \onlinecite{else21}, which characterizes a particular point on the Fermi surface.  However, unlike that description, $\chi_F$ provides a global characterization of the Fermi surface.
For $D=1$ the bipartite entanglement entropy has a universal term $S = (c/3){\rm log} L$\cite{wilczek94,calabrese04} with $c=\chi_F$.   This has been generalized to higher $D$, where $S$ describes a logarithmic area law entanglement that involves the projected area of the Fermi surface\cite{gioev06,swingle10}, which is non-zero even for a system of decoupled 1D wires\cite{ding12}, with $\chi_F=0$.   We speculate that for $D>1$, $\chi_F$ shows up in an intrinsically $D$ dimensional entanglement measure.

We thank Patrick Lee and Pok Man Tam
for helpful suggestions.  This work was supported by a Simons Investigator Grant from the Simons Foundation.


\begin{thebibliography}{42}%
\makeatletter
\providecommand \@ifxundefined [1]{%
 \@ifx{#1\undefined}
}%
\providecommand \@ifnum [1]{%
 \ifnum #1\expandafter \@firstoftwo
 \else \expandafter \@secondoftwo
 \fi
}%
\providecommand \@ifx [1]{%
 \ifx #1\expandafter \@firstoftwo
 \else \expandafter \@secondoftwo
 \fi
}%
\providecommand \natexlab [1]{#1}%
\providecommand \enquote  [1]{``#1''}%
\providecommand \bibnamefont  [1]{#1}%
\providecommand \bibfnamefont [1]{#1}%
\providecommand \citenamefont [1]{#1}%
\providecommand \href@noop [0]{\@secondoftwo}%
\providecommand \href [0]{\begingroup \@sanitize@url \@href}%
\providecommand \@href[1]{\@@startlink{#1}\@@href}%
\providecommand \@@href[1]{\endgroup#1\@@endlink}%
\providecommand \@sanitize@url [0]{\catcode `\\12\catcode `\$12\catcode
  `\&12\catcode `\#12\catcode `\^12\catcode `\_12\catcode `\%12\relax}%
\providecommand \@@startlink[1]{}%
\providecommand \@@endlink[0]{}%
\providecommand \url  [0]{\begingroup\@sanitize@url \@url }%
\providecommand \@url [1]{\endgroup\@href {#1}{\urlprefix }}%
\providecommand \urlprefix  [0]{URL }%
\providecommand \Eprint [0]{\href }%
\providecommand \doibase [0]{http://dx.doi.org/}%
\providecommand \selectlanguage [0]{\@gobble}%
\providecommand \bibinfo  [0]{\@secondoftwo}%
\providecommand \bibfield  [0]{\@secondoftwo}%
\providecommand \translation [1]{[#1]}%
\providecommand \BibitemOpen [0]{}%
\providecommand \bibitemStop [0]{}%
\providecommand \bibitemNoStop [0]{.\EOS\space}%
\providecommand \EOS [0]{\spacefactor3000\relax}%
\providecommand \BibitemShut  [1]{\csname bibitem#1\endcsname}%
\let\auto@bib@innerbib\@empty
\bibitem [{\citenamefont {Klitzing}\ \emph {et~al.}(1980)\citenamefont
  {Klitzing}, \citenamefont {Dorda},\ and\ \citenamefont
  {Pepper}}]{vonklitzing80}%
  \BibitemOpen
  \bibfield  {author} {\bibinfo {author} {\bibfnamefont {K.~v.}\ \bibnamefont
  {Klitzing}}, \bibinfo {author} {\bibfnamefont {G.}~\bibnamefont {Dorda}}, \
  and\ \bibinfo {author} {\bibfnamefont {M.}~\bibnamefont {Pepper}},\
  }\bibfield  {title} {\enquote {\bibinfo {title} {New method for high-accuracy
  determination of the fine-structure constant based on quantized hall
  resistance},}\ }\href {\doibase 10.1103/PhysRevLett.45.494} {\bibfield
  {journal} {\bibinfo  {journal} {Phys. Rev. Lett.}\ }\textbf {\bibinfo
  {volume} {45}},\ \bibinfo {pages} {494--497} (\bibinfo {year}
  {1980})}\BibitemShut {NoStop}%
\bibitem [{\citenamefont {Thouless}\ \emph {et~al.}(1982)\citenamefont
  {Thouless}, \citenamefont {Kohmoto}, \citenamefont {Nightingale},\ and\
  \citenamefont {den Nijs}}]{tknn82}%
  \BibitemOpen
  \bibfield  {author} {\bibinfo {author} {\bibfnamefont {D.~J.}\ \bibnamefont
  {Thouless}}, \bibinfo {author} {\bibfnamefont {M.}~\bibnamefont {Kohmoto}},
  \bibinfo {author} {\bibfnamefont {M.~P.}\ \bibnamefont {Nightingale}}, \ and\
  \bibinfo {author} {\bibfnamefont {M.}~\bibnamefont {den Nijs}},\ }\bibfield
  {title} {\enquote {\bibinfo {title} {Quantized hall conductance in a
  two-dimensional periodic potential},}\ }\href {\doibase
  10.1103/PhysRevLett.49.405} {\bibfield  {journal} {\bibinfo  {journal} {Phys.
  Rev. Lett.}\ }\textbf {\bibinfo {volume} {49}},\ \bibinfo {pages} {405--408}
  (\bibinfo {year} {1982})}\BibitemShut {NoStop}%
\bibitem [{\citenamefont {Haldane}(2004)}]{haldane04}%
  \BibitemOpen
  \bibfield  {author} {\bibinfo {author} {\bibfnamefont {F.~D.~M.}\
  \bibnamefont {Haldane}},\ }\bibfield  {title} {\enquote {\bibinfo {title}
  {Berry curvature on the fermi surface: Anomalous hall effect as a topological
  fermi-liquid property},}\ }\href {\doibase 10.1103/PhysRevLett.93.206602}
  {\bibfield  {journal} {\bibinfo  {journal} {Phys. Rev. Lett.}\ }\textbf
  {\bibinfo {volume} {93}},\ \bibinfo {pages} {206602} (\bibinfo {year}
  {2004})}\BibitemShut {NoStop}%
\bibitem [{\citenamefont {de~Juan}\ \emph {et~al.}(2017)\citenamefont
  {de~Juan}, \citenamefont {Grushin}, \citenamefont {Morimoto},\ and\
  \citenamefont {Moore}}]{dejuan17}%
  \BibitemOpen
  \bibfield  {author} {\bibinfo {author} {\bibfnamefont {F.}~\bibnamefont
  {de~Juan}}, \bibinfo {author} {\bibfnamefont {A.G.}\ \bibnamefont {Grushin}},
  \bibinfo {author} {\bibfnamefont {T.}~\bibnamefont {Morimoto}}, \ and\
  \bibinfo {author} {\bibfnamefont {J.~E.}\ \bibnamefont {Moore}},\ }\bibfield
  {title} {\enquote {\bibinfo {title} {Quantized circular photogalvanic effect
  in weyl semimetals},}\ }\href {\doibase 10.1038/ncomms15995} {\bibfield
  {journal} {\bibinfo  {journal} {Nature Communications}\ }\textbf {\bibinfo
  {volume} {8}},\ \bibinfo {pages} {15995} (\bibinfo {year}
  {2017})}\BibitemShut {NoStop}%
\bibitem [{\citenamefont {Avdoshkin}\ \emph {et~al.}(2020)\citenamefont
  {Avdoshkin}, \citenamefont {Kozii},\ and\ \citenamefont
  {Moore}}]{avdoshkin20}%
  \BibitemOpen
  \bibfield  {author} {\bibinfo {author} {\bibfnamefont {Alexander}\
  \bibnamefont {Avdoshkin}}, \bibinfo {author} {\bibfnamefont {Vladyslav}\
  \bibnamefont {Kozii}}, \ and\ \bibinfo {author} {\bibfnamefont {Joel~E.}\
  \bibnamefont {Moore}},\ }\bibfield  {title} {\enquote {\bibinfo {title}
  {Interactions remove the quantization of the chiral photocurrent at weyl
  points},}\ }\href {\doibase 10.1103/PhysRevLett.124.196603} {\bibfield
  {journal} {\bibinfo  {journal} {Phys. Rev. Lett.}\ }\textbf {\bibinfo
  {volume} {124}},\ \bibinfo {pages} {196603} (\bibinfo {year}
  {2020})}\BibitemShut {NoStop}%
\bibitem [{\citenamefont {Ashcroft}\ and\ \citenamefont
  {Mermin}(1976)}]{ashcroft76}%
  \BibitemOpen
  \bibfield  {author} {\bibinfo {author} {\bibfnamefont {N.~W.}\ \bibnamefont
  {Ashcroft}}\ and\ \bibinfo {author} {\bibfnamefont {N.~D.}\ \bibnamefont
  {Mermin}},\ }\href@noop {} {\emph {\bibinfo {title} {{S}olid {S}tate
  {P}hysics}}}\ (\bibinfo  {publisher} {Holt-Saunders},\ \bibinfo {year}
  {1976})\BibitemShut {NoStop}%
\bibitem [{\citenamefont {Lifshitz}(1960)}]{lifshitz60}%
  \BibitemOpen
  \bibfield  {author} {\bibinfo {author} {\bibfnamefont {I.~M.}\ \bibnamefont
  {Lifshitz}},\ }\bibfield  {title} {\enquote {\bibinfo {title} {Anomalies of
  electron characteristics of a metal in the high pressure region},}\
  }\href@noop {} {\bibfield  {journal} {\bibinfo  {journal} {Soviet Physics
  JETP}\ }\textbf {\bibinfo {volume} {11}},\ \bibinfo {pages} {1130--1135}
  (\bibinfo {year} {1960})}\BibitemShut {NoStop}%
\bibitem [{\citenamefont {Volovik}(2017)}]{volovik17}%
  \BibitemOpen
  \bibfield  {author} {\bibinfo {author} {\bibfnamefont {G~E}\ \bibnamefont
  {Volovik}},\ }\bibfield  {title} {\enquote {\bibinfo {title} {Topological
  lifshitz transitions},}\ }\href {\doibase 10.1063/1.4974185} {\bibfield
  {journal} {\bibinfo  {journal} {Low Temperature Physics}\ }\textbf {\bibinfo
  {volume} {43}},\ \bibinfo {pages} {47--55} (\bibinfo {year}
  {2017})}\BibitemShut {NoStop}%
\bibitem [{\citenamefont {Volovik}(2018)}]{volovik18}%
  \BibitemOpen
  \bibfield  {author} {\bibinfo {author} {\bibfnamefont {G~E}\ \bibnamefont
  {Volovik}},\ }\bibfield  {title} {\enquote {\bibinfo {title} {Exotic lifshitz
  transitions in topological materials},}\ }\href {\doibase
  10.3367/ufne.2017.01.038218} {\bibfield  {journal} {\bibinfo  {journal}
  {Physics-Uspekhi}\ }\textbf {\bibinfo {volume} {61}},\ \bibinfo {pages}
  {89--98} (\bibinfo {year} {2018})}\BibitemShut {NoStop}%
\bibitem [{\citenamefont {Landauer}(1957)}]{landauer57}%
  \BibitemOpen
  \bibfield  {author} {\bibinfo {author} {\bibfnamefont {R.}~\bibnamefont
  {Landauer}},\ }\bibfield  {title} {\enquote {\bibinfo {title} {Spatial
  variation of currents and fields due to localized scatterers in metallic
  conduction},}\ }\href {\doibase 10.1147/rd.13.0223} {\bibfield  {journal}
  {\bibinfo  {journal} {IBM Journal of Research and Development}\ }\textbf
  {\bibinfo {volume} {1}},\ \bibinfo {pages} {223--231} (\bibinfo {year}
  {1957})}\BibitemShut {NoStop}%
\bibitem [{\citenamefont {Fisher}\ and\ \citenamefont {Lee}(1981)}]{fisher81}%
  \BibitemOpen
  \bibfield  {author} {\bibinfo {author} {\bibfnamefont {Daniel~S.}\
  \bibnamefont {Fisher}}\ and\ \bibinfo {author} {\bibfnamefont {Patrick~A.}\
  \bibnamefont {Lee}},\ }\bibfield  {title} {\enquote {\bibinfo {title}
  {Relation between conductivity and transmission matrix},}\ }\href {\doibase
  10.1103/PhysRevB.23.6851} {\bibfield  {journal} {\bibinfo  {journal} {Phys.
  Rev. B}\ }\textbf {\bibinfo {volume} {23}},\ \bibinfo {pages} {6851--6854}
  (\bibinfo {year} {1981})}\BibitemShut {NoStop}%
\bibitem [{\citenamefont {van Wees}\ \emph {et~al.}(1988)\citenamefont {van
  Wees}, \citenamefont {van Houten}, \citenamefont {Beenakker}, \citenamefont
  {Williamson}, \citenamefont {Kouwenhoven}, \citenamefont {van~der Marel},\
  and\ \citenamefont {Foxon}}]{vanwees88}%
  \BibitemOpen
  \bibfield  {author} {\bibinfo {author} {\bibfnamefont {B.~J.}\ \bibnamefont
  {van Wees}}, \bibinfo {author} {\bibfnamefont {H.}~\bibnamefont {van
  Houten}}, \bibinfo {author} {\bibfnamefont {C.~W.~J.}\ \bibnamefont
  {Beenakker}}, \bibinfo {author} {\bibfnamefont {J.~G.}\ \bibnamefont
  {Williamson}}, \bibinfo {author} {\bibfnamefont {L.~P.}\ \bibnamefont
  {Kouwenhoven}}, \bibinfo {author} {\bibfnamefont {D.}~\bibnamefont {van~der
  Marel}}, \ and\ \bibinfo {author} {\bibfnamefont {C.~T.}\ \bibnamefont
  {Foxon}},\ }\bibfield  {title} {\enquote {\bibinfo {title} {Quantized
  conductance of point contacts in a two-dimensional electron gas},}\ }\href
  {\doibase 10.1103/PhysRevLett.60.848} {\bibfield  {journal} {\bibinfo
  {journal} {Phys. Rev. Lett.}\ }\textbf {\bibinfo {volume} {60}},\ \bibinfo
  {pages} {848--850} (\bibinfo {year} {1988})}\BibitemShut {NoStop}%
\bibitem [{\citenamefont {Honda}\ \emph {et~al.}(1995)\citenamefont {Honda},
  \citenamefont {Tarucha}, \citenamefont {Saku},\ and\ \citenamefont
  {Tokura}}]{honda95}%
  \BibitemOpen
  \bibfield  {author} {\bibinfo {author} {\bibfnamefont {Takashi}\ \bibnamefont
  {Honda}}, \bibinfo {author} {\bibfnamefont {Seigo}\ \bibnamefont {Tarucha}},
  \bibinfo {author} {\bibfnamefont {Tadashi}\ \bibnamefont {Saku}}, \ and\
  \bibinfo {author} {\bibfnamefont {Yasuhiro}\ \bibnamefont {Tokura}},\
  }\bibfield  {title} {\enquote {\bibinfo {title} {Quantized conductance
  observed in quantum wires 2 to 10 $\mathrm{\mu}$ m long},}\ }\href {\doibase
  10.1143/jjap.34.l72} {\bibfield  {journal} {\bibinfo  {journal} {Japanese
  Journal of Applied Physics}\ }\textbf {\bibinfo {volume} {34}},\ \bibinfo
  {pages} {L72--L75} (\bibinfo {year} {1995})}\BibitemShut {NoStop}%
\bibitem [{\citenamefont {van Weperen}\ \emph {et~al.}(2013)\citenamefont {van
  Weperen}, \citenamefont {Plissard}, \citenamefont {Bakkers}, \citenamefont
  {Frolov},\ and\ \citenamefont {Kouwenhoven}}]{vanweperen13}%
  \BibitemOpen
  \bibfield  {author} {\bibinfo {author} {\bibfnamefont {Ilse}\ \bibnamefont
  {van Weperen}}, \bibinfo {author} {\bibfnamefont {Sébastien~R.}\
  \bibnamefont {Plissard}}, \bibinfo {author} {\bibfnamefont {Erik P. A.~M.}\
  \bibnamefont {Bakkers}}, \bibinfo {author} {\bibfnamefont {Sergey~M.}\
  \bibnamefont {Frolov}}, \ and\ \bibinfo {author} {\bibfnamefont {Leo~P.}\
  \bibnamefont {Kouwenhoven}},\ }\bibfield  {title} {\enquote {\bibinfo {title}
  {Quantized conductance in an insb nanowire},}\ }\href {\doibase
  10.1021/nl3035256} {\bibfield  {journal} {\bibinfo  {journal} {Nano Letters}\
  }\textbf {\bibinfo {volume} {13}},\ \bibinfo {pages} {387--391} (\bibinfo
  {year} {2013})}\BibitemShut {NoStop}%
\bibitem [{\citenamefont {Frank}\ \emph {et~al.}(1998)\citenamefont {Frank},
  \citenamefont {Poncharal}, \citenamefont {Wang},\ and\ \citenamefont
  {de~Heer}}]{frank98}%
  \BibitemOpen
  \bibfield  {author} {\bibinfo {author} {\bibfnamefont {Stefan}\ \bibnamefont
  {Frank}}, \bibinfo {author} {\bibfnamefont {Philippe}\ \bibnamefont
  {Poncharal}}, \bibinfo {author} {\bibfnamefont {Z.~L.}\ \bibnamefont {Wang}},
  \ and\ \bibinfo {author} {\bibfnamefont {Walt~A.}\ \bibnamefont {de~Heer}},\
  }\bibfield  {title} {\enquote {\bibinfo {title} {Carbon nanotube quantum
  resistors},}\ }\href {\doibase 10.1126/science.280.5370.1744} {\bibfield
  {journal} {\bibinfo  {journal} {Science}\ }\textbf {\bibinfo {volume}
  {280}},\ \bibinfo {pages} {1744--1746} (\bibinfo {year} {1998})}\BibitemShut
  {NoStop}%
\bibitem [{\citenamefont {Nakahara}(1990)}]{nakahara90}%
  \BibitemOpen
  \bibfield  {author} {\bibinfo {author} {\bibfnamefont {Mikio}\ \bibnamefont
  {Nakahara}},\ }\href {https://cds.cern.ch/record/206619} {\emph {\bibinfo
  {title} {{Geometry, topology and physics}}}},\ Graduate student series in
  physics\ (\bibinfo  {publisher} {Hilger},\ \bibinfo {address} {Bristol},\
  \bibinfo {year} {1990})\BibitemShut {NoStop}%
\bibitem [{\citenamefont {Milnor}(1969)}]{milnor69}%
  \BibitemOpen
  \bibfield  {author} {\bibinfo {author} {\bibfnamefont {J.}~\bibnamefont
  {Milnor}},\ }\href {http://www.jstor.org/stable/j.ctv3f8rb6} {\emph {\bibinfo
  {title} {Morse Theory. (AM-51), Volume 51}}}\ (\bibinfo  {publisher}
  {Princeton University Press},\ \bibinfo {year} {1969})\BibitemShut {NoStop}%
\bibitem [{\citenamefont {Nash}\ and\ \citenamefont {Sen}(1988)}]{nash88}%
  \BibitemOpen
  \bibfield  {author} {\bibinfo {author} {\bibfnamefont {Charles}\ \bibnamefont
  {Nash}}\ and\ \bibinfo {author} {\bibfnamefont {Siddhartha}\ \bibnamefont
  {Sen}},\ }\href
  {http://www.amazon.com/Topology-Geometry-Physicists-Charles-Nash/dp/0125140819/ref=sr_1_1?ie=UTF8&s=books&qid=1263990064&sr=1-1}
  {\emph {\bibinfo {title} {Topology and Geometry for Physicists}}}\ (\bibinfo
  {publisher} {Academic Press},\ \bibinfo {year} {1988})\BibitemShut {NoStop}%
\bibitem [{\citenamefont {Laughlin}(1981)}]{laughlin81}%
  \BibitemOpen
  \bibfield  {author} {\bibinfo {author} {\bibfnamefont {R.~B.}\ \bibnamefont
  {Laughlin}},\ }\bibfield  {title} {\enquote {\bibinfo {title} {Quantized hall
  conductivity in two dimensions},}\ }\href {\doibase 10.1103/PhysRevB.23.5632}
  {\bibfield  {journal} {\bibinfo  {journal} {Phys. Rev. B}\ }\textbf {\bibinfo
  {volume} {23}},\ \bibinfo {pages} {5632--5633} (\bibinfo {year}
  {1981})}\BibitemShut {NoStop}%
\bibitem [{\citenamefont {Kraut}\ and\ \citenamefont {von
  Baltz}(1979)}]{kraut79}%
  \BibitemOpen
  \bibfield  {author} {\bibinfo {author} {\bibfnamefont {Wolfgang}\
  \bibnamefont {Kraut}}\ and\ \bibinfo {author} {\bibfnamefont {Ralph}\
  \bibnamefont {von Baltz}},\ }\bibfield  {title} {\enquote {\bibinfo {title}
  {Anomalous bulk photovoltaic effect in ferroelectrics: A quadratic response
  theory},}\ }\href {\doibase 10.1103/PhysRevB.19.1548} {\bibfield  {journal}
  {\bibinfo  {journal} {Phys. Rev. B}\ }\textbf {\bibinfo {volume} {19}},\
  \bibinfo {pages} {1548--1554} (\bibinfo {year} {1979})}\BibitemShut {NoStop}%
\bibitem [{\citenamefont {von Baltz}\ and\ \citenamefont
  {Kraut}(1981)}]{vonbaltz81}%
  \BibitemOpen
  \bibfield  {author} {\bibinfo {author} {\bibfnamefont {Ralph}\ \bibnamefont
  {von Baltz}}\ and\ \bibinfo {author} {\bibfnamefont {Wolfgang}\ \bibnamefont
  {Kraut}},\ }\bibfield  {title} {\enquote {\bibinfo {title} {Theory of the
  bulk photovoltaic effect in pure crystals},}\ }\href {\doibase
  10.1103/PhysRevB.23.5590} {\bibfield  {journal} {\bibinfo  {journal} {Phys.
  Rev. B}\ }\textbf {\bibinfo {volume} {23}},\ \bibinfo {pages} {5590--5596}
  (\bibinfo {year} {1981})}\BibitemShut {NoStop}%
\bibitem [{\citenamefont {Zhang}\ \emph {et~al.}(2018)\citenamefont {Zhang},
  \citenamefont {Ishizuka}, \citenamefont {van~den Brink}, \citenamefont
  {Felser}, \citenamefont {Yan},\ and\ \citenamefont {Nagaosa}}]{zhang18}%
  \BibitemOpen
  \bibfield  {author} {\bibinfo {author} {\bibfnamefont {Yang}\ \bibnamefont
  {Zhang}}, \bibinfo {author} {\bibfnamefont {Hiroaki}\ \bibnamefont
  {Ishizuka}}, \bibinfo {author} {\bibfnamefont {Jeroen}\ \bibnamefont {van~den
  Brink}}, \bibinfo {author} {\bibfnamefont {Claudia}\ \bibnamefont {Felser}},
  \bibinfo {author} {\bibfnamefont {Binghai}\ \bibnamefont {Yan}}, \ and\
  \bibinfo {author} {\bibfnamefont {Naoto}\ \bibnamefont {Nagaosa}},\
  }\bibfield  {title} {\enquote {\bibinfo {title} {Photogalvanic effect in weyl
  semimetals from first principles},}\ }\href {\doibase
  10.1103/PhysRevB.97.241118} {\bibfield  {journal} {\bibinfo  {journal} {Phys.
  Rev. B}\ }\textbf {\bibinfo {volume} {97}},\ \bibinfo {pages} {241118}
  (\bibinfo {year} {2018})}\BibitemShut {NoStop}%
\bibitem [{Note1()}]{Note1}%
  \BibitemOpen
  \bibinfo {note} {If one of the angles $\varphi _p$ is greater than $\pi $,
  $\alpha _i$ is still quantized, but its value is modified. See supplemental
  section \ref {Appendix B}, where it is also established that for Fig. 1a the
  pulse argument analysis remains valid.}\BibitemShut {Stop}%
\bibitem [{\citenamefont {Apel}\ and\ \citenamefont {Rice}(1982)}]{apel82}%
  \BibitemOpen
  \bibfield  {author} {\bibinfo {author} {\bibfnamefont {W.}~\bibnamefont
  {Apel}}\ and\ \bibinfo {author} {\bibfnamefont {T.~M.}\ \bibnamefont
  {Rice}},\ }\bibfield  {title} {\enquote {\bibinfo {title} {Combined effect of
  disorder and interaction on the conductance of a one-dimensional fermion
  system},}\ }\href {\doibase 10.1103/PhysRevB.26.7063} {\bibfield  {journal}
  {\bibinfo  {journal} {Phys. Rev. B}\ }\textbf {\bibinfo {volume} {26}},\
  \bibinfo {pages} {7063--7065} (\bibinfo {year} {1982})}\BibitemShut {NoStop}%
\bibitem [{\citenamefont {Kane}\ and\ \citenamefont {Fisher}(1992)}]{kane92}%
  \BibitemOpen
  \bibfield  {author} {\bibinfo {author} {\bibfnamefont {C.~L.}\ \bibnamefont
  {Kane}}\ and\ \bibinfo {author} {\bibfnamefont {Matthew P.~A.}\ \bibnamefont
  {Fisher}},\ }\bibfield  {title} {\enquote {\bibinfo {title} {Transport in a
  one-channel luttinger liquid},}\ }\href {\doibase
  10.1103/PhysRevLett.68.1220} {\bibfield  {journal} {\bibinfo  {journal}
  {Phys. Rev. Lett.}\ }\textbf {\bibinfo {volume} {68}},\ \bibinfo {pages}
  {1220--1223} (\bibinfo {year} {1992})}\BibitemShut {NoStop}%
\bibitem [{\citenamefont {Maslov}\ and\ \citenamefont
  {Stone}(1995)}]{maslov95}%
  \BibitemOpen
  \bibfield  {author} {\bibinfo {author} {\bibfnamefont {Dmitrii~L.}\
  \bibnamefont {Maslov}}\ and\ \bibinfo {author} {\bibfnamefont {Michael}\
  \bibnamefont {Stone}},\ }\bibfield  {title} {\enquote {\bibinfo {title}
  {Landauer conductance of luttinger liquids with leads},}\ }\href {\doibase
  10.1103/PhysRevB.52.R5539} {\bibfield  {journal} {\bibinfo  {journal} {Phys.
  Rev. B}\ }\textbf {\bibinfo {volume} {52}},\ \bibinfo {pages} {R5539--R5542}
  (\bibinfo {year} {1995})}\BibitemShut {NoStop}%
\bibitem [{\citenamefont {Ponomarenko}(1995)}]{ponomarenko95}%
  \BibitemOpen
  \bibfield  {author} {\bibinfo {author} {\bibfnamefont {V.~V.}\ \bibnamefont
  {Ponomarenko}},\ }\bibfield  {title} {\enquote {\bibinfo {title}
  {Renormalization of the one-dimensional conductance in the luttinger-liquid
  model},}\ }\href {\doibase 10.1103/PhysRevB.52.R8666} {\bibfield  {journal}
  {\bibinfo  {journal} {Phys. Rev. B}\ }\textbf {\bibinfo {volume} {52}},\
  \bibinfo {pages} {R8666--R8667} (\bibinfo {year} {1995})}\BibitemShut
  {NoStop}%
\bibitem [{\citenamefont {Safi}\ and\ \citenamefont {Schulz}(1995)}]{safi95}%
  \BibitemOpen
  \bibfield  {author} {\bibinfo {author} {\bibfnamefont {I.}~\bibnamefont
  {Safi}}\ and\ \bibinfo {author} {\bibfnamefont {H.~J.}\ \bibnamefont
  {Schulz}},\ }\bibfield  {title} {\enquote {\bibinfo {title} {Transport in an
  inhomogeneous interacting one-dimensional system},}\ }\href {\doibase
  10.1103/PhysRevB.52.R17040} {\bibfield  {journal} {\bibinfo  {journal} {Phys.
  Rev. B}\ }\textbf {\bibinfo {volume} {52}},\ \bibinfo {pages}
  {R17040--R17043} (\bibinfo {year} {1995})}\BibitemShut {NoStop}%
\bibitem [{\citenamefont {Safi}(1997)}]{safi97}%
  \BibitemOpen
  \bibfield  {author} {\bibinfo {author} {\bibfnamefont {In`es}\ \bibnamefont
  {Safi}},\ }\bibfield  {title} {\enquote {\bibinfo {title} {Conductance of a
  quantum wire: Landauer's approach versus the kubo formula},}\ }\href
  {\doibase 10.1103/PhysRevB.55.R7331} {\bibfield  {journal} {\bibinfo
  {journal} {Phys. Rev. B}\ }\textbf {\bibinfo {volume} {55}},\ \bibinfo
  {pages} {R7331--R7334} (\bibinfo {year} {1997})}\BibitemShut {NoStop}%
\bibitem [{\citenamefont {Safi}(1999)}]{safi99}%
  \BibitemOpen
  \bibfield  {author} {\bibinfo {author} {\bibfnamefont {I.}~\bibnamefont
  {Safi}},\ }\bibfield  {title} {\enquote {\bibinfo {title} {A dynamic
  scattering approach for a gated interacting wire},}\ }\href {\doibase
  10.1007/s100510051026} {\bibfield  {journal} {\bibinfo  {journal} {European
  Physical Journal B}\ }\textbf {\bibinfo {volume} {12}},\ \bibinfo {pages}
  {451} (\bibinfo {year} {1999})}\BibitemShut {NoStop}%
\bibitem [{\citenamefont {Thomale}\ and\ \citenamefont
  {Seidel}(2011)}]{thomale11}%
  \BibitemOpen
  \bibfield  {author} {\bibinfo {author} {\bibfnamefont {Ronny}\ \bibnamefont
  {Thomale}}\ and\ \bibinfo {author} {\bibfnamefont {Alexander}\ \bibnamefont
  {Seidel}},\ }\bibfield  {title} {\enquote {\bibinfo {title} {Minimal model of
  quantized conductance in interacting ballistic quantum wires},}\ }\href
  {\doibase 10.1103/PhysRevB.83.115330} {\bibfield  {journal} {\bibinfo
  {journal} {Phys. Rev. B}\ }\textbf {\bibinfo {volume} {83}},\ \bibinfo
  {pages} {115330} (\bibinfo {year} {2011})}\BibitemShut {NoStop}%
\bibitem [{\citenamefont {Kawabata}(1996)}]{kawabata95}%
  \BibitemOpen
  \bibfield  {author} {\bibinfo {author} {\bibfnamefont {Arisato}\ \bibnamefont
  {Kawabata}},\ }\bibfield  {title} {\enquote {\bibinfo {title} {On the
  renormalization of conductance in tomonaga-luttinger liquid},}\ }\href
  {\doibase 10.1143/JPSJ.65.30} {\bibfield  {journal} {\bibinfo  {journal}
  {Journal of the Physical Society of Japan}\ }\textbf {\bibinfo {volume}
  {65}},\ \bibinfo {pages} {30--32} (\bibinfo {year} {1996})}\BibitemShut
  {NoStop}%
\bibitem [{\citenamefont {Shimizu}(1996)}]{shimizu96}%
  \BibitemOpen
  \bibfield  {author} {\bibinfo {author} {\bibfnamefont {Akira}\ \bibnamefont
  {Shimizu}},\ }\bibfield  {title} {\enquote {\bibinfo {title} {Landauer
  conductance and nonequilibrium noise of one-dimensional interacting electron
  systems},}\ }\href {\doibase 10.1143/JPSJ.65.1162} {\bibfield  {journal}
  {\bibinfo  {journal} {Journal of the Physical Society of Japan}\ }\textbf
  {\bibinfo {volume} {65}},\ \bibinfo {pages} {1162--1165} (\bibinfo {year}
  {1996})}\BibitemShut {NoStop}%
\bibitem [{\citenamefont {Pines}\ and\ \citenamefont
  {Nozieres}(1966)}]{pines66}%
  \BibitemOpen
  \bibfield  {author} {\bibinfo {author} {\bibfnamefont {David}\ \bibnamefont
  {Pines}}\ and\ \bibinfo {author} {\bibfnamefont {Philippe}\ \bibnamefont
  {Nozieres}},\ }\href {\doibase https://doi.org/10.4324/9780429492662} {\emph
  {\bibinfo {title} {The Theory of Quantum Liquids}}}\ (\bibinfo  {publisher}
  {CRC Press},\ \bibinfo {address} {Boca Raton},\ \bibinfo {year}
  {1966})\BibitemShut {NoStop}%
\bibitem [{\citenamefont {Shankar}(1994)}]{shankar94}%
  \BibitemOpen
  \bibfield  {author} {\bibinfo {author} {\bibfnamefont {R.}~\bibnamefont
  {Shankar}},\ }\bibfield  {title} {\enquote {\bibinfo {title}
  {Renormalization-group approach to interacting fermions},}\ }\href {\doibase
  10.1103/RevModPhys.66.129} {\bibfield  {journal} {\bibinfo  {journal} {Rev.
  Mod. Phys.}\ }\textbf {\bibinfo {volume} {66}},\ \bibinfo {pages} {129--192}
  (\bibinfo {year} {1994})}\BibitemShut {NoStop}%
\bibitem [{\citenamefont {Rostami}\ \emph {et~al.}(2017)\citenamefont
  {Rostami}, \citenamefont {Katsnelson},\ and\ \citenamefont
  {Polini}}]{rostami17}%
  \BibitemOpen
  \bibfield  {author} {\bibinfo {author} {\bibfnamefont {Habib}\ \bibnamefont
  {Rostami}}, \bibinfo {author} {\bibfnamefont {Mikhail~I.}\ \bibnamefont
  {Katsnelson}}, \ and\ \bibinfo {author} {\bibfnamefont {Marco}\ \bibnamefont
  {Polini}},\ }\bibfield  {title} {\enquote {\bibinfo {title} {Theory of
  plasmonic effects in nonlinear optics: The case of graphene},}\ }\href
  {\doibase 10.1103/PhysRevB.95.035416} {\bibfield  {journal} {\bibinfo
  {journal} {Phys. Rev. B}\ }\textbf {\bibinfo {volume} {95}},\ \bibinfo
  {pages} {035416} (\bibinfo {year} {2017})}\BibitemShut {NoStop}%
\bibitem [{\citenamefont {Else}\ \emph {et~al.}(2021)\citenamefont {Else},
  \citenamefont {Thorngren},\ and\ \citenamefont {Senthil}}]{else21}%
  \BibitemOpen
  \bibfield  {author} {\bibinfo {author} {\bibfnamefont {Dominic~V.}\
  \bibnamefont {Else}}, \bibinfo {author} {\bibfnamefont {Ryan}\ \bibnamefont
  {Thorngren}}, \ and\ \bibinfo {author} {\bibfnamefont {T.}~\bibnamefont
  {Senthil}},\ }\bibfield  {title} {\enquote {\bibinfo {title} {Non-fermi
  liquids as ersatz fermi liquids: General constraints on compressible
  metals},}\ }\href {\doibase 10.1103/PhysRevX.11.021005} {\bibfield  {journal}
  {\bibinfo  {journal} {Phys. Rev. X}\ }\textbf {\bibinfo {volume} {11}},\
  \bibinfo {pages} {021005} (\bibinfo {year} {2021})}\BibitemShut {NoStop}%
\bibitem [{\citenamefont {Holzhey}\ \emph {et~al.}(1994)\citenamefont
  {Holzhey}, \citenamefont {Larsen},\ and\ \citenamefont
  {Wilczek}}]{wilczek94}%
  \BibitemOpen
  \bibfield  {author} {\bibinfo {author} {\bibfnamefont {Christoph}\
  \bibnamefont {Holzhey}}, \bibinfo {author} {\bibfnamefont {Finn}\
  \bibnamefont {Larsen}}, \ and\ \bibinfo {author} {\bibfnamefont {Frank}\
  \bibnamefont {Wilczek}},\ }\bibfield  {title} {\enquote {\bibinfo {title}
  {Geometric and renormalized entropy in conformal field theory},}\ }\href
  {\doibase https://doi.org/10.1016/0550-3213(94)90402-2} {\bibfield  {journal}
  {\bibinfo  {journal} {Nuclear Physics B}\ }\textbf {\bibinfo {volume}
  {424}},\ \bibinfo {pages} {443--467} (\bibinfo {year} {1994})}\BibitemShut
  {NoStop}%
\bibitem [{\citenamefont {Calabrese}\ and\ \citenamefont
  {Cardy}(2004)}]{calabrese04}%
  \BibitemOpen
  \bibfield  {author} {\bibinfo {author} {\bibfnamefont {Pasquale}\
  \bibnamefont {Calabrese}}\ and\ \bibinfo {author} {\bibfnamefont {John}\
  \bibnamefont {Cardy}},\ }\bibfield  {title} {\enquote {\bibinfo {title}
  {Entanglement entropy and quantum field theory},}\ }\href {\doibase
  10.1088/1742-5468/2004/06/p06002} {\bibfield  {journal} {\bibinfo  {journal}
  {Journal of Statistical Mechanics: Theory and Experiment}\ }\textbf {\bibinfo
  {volume} {2004}},\ \bibinfo {pages} {P06002} (\bibinfo {year}
  {2004})}\BibitemShut {NoStop}%
\bibitem [{\citenamefont {Gioev}\ and\ \citenamefont {Klich}(2006)}]{gioev06}%
  \BibitemOpen
  \bibfield  {author} {\bibinfo {author} {\bibfnamefont {Dimitri}\ \bibnamefont
  {Gioev}}\ and\ \bibinfo {author} {\bibfnamefont {Israel}\ \bibnamefont
  {Klich}},\ }\bibfield  {title} {\enquote {\bibinfo {title} {Entanglement
  entropy of fermions in any dimension and the widom conjecture},}\ }\href
  {\doibase 10.1103/PhysRevLett.96.100503} {\bibfield  {journal} {\bibinfo
  {journal} {Phys. Rev. Lett.}\ }\textbf {\bibinfo {volume} {96}},\ \bibinfo
  {pages} {100503} (\bibinfo {year} {2006})}\BibitemShut {NoStop}%
\bibitem [{\citenamefont {Swingle}(2010)}]{swingle10}%
  \BibitemOpen
  \bibfield  {author} {\bibinfo {author} {\bibfnamefont {Brian}\ \bibnamefont
  {Swingle}},\ }\bibfield  {title} {\enquote {\bibinfo {title} {Entanglement
  entropy and the fermi surface},}\ }\href {\doibase
  10.1103/PhysRevLett.105.050502} {\bibfield  {journal} {\bibinfo  {journal}
  {Phys. Rev. Lett.}\ }\textbf {\bibinfo {volume} {105}},\ \bibinfo {pages}
  {050502} (\bibinfo {year} {2010})}\BibitemShut {NoStop}%
\bibitem [{\citenamefont {Ding}\ \emph {et~al.}(2012)\citenamefont {Ding},
  \citenamefont {Seidel},\ and\ \citenamefont {Yang}}]{ding12}%
  \BibitemOpen
  \bibfield  {author} {\bibinfo {author} {\bibfnamefont {Wenxin}\ \bibnamefont
  {Ding}}, \bibinfo {author} {\bibfnamefont {Alexander}\ \bibnamefont
  {Seidel}}, \ and\ \bibinfo {author} {\bibfnamefont {Kun}\ \bibnamefont
  {Yang}},\ }\bibfield  {title} {\enquote {\bibinfo {title} {Entanglement
  entropy of fermi liquids via multidimensional bosonization},}\ }\href
  {\doibase 10.1103/PhysRevX.2.011012} {\bibfield  {journal} {\bibinfo
  {journal} {Phys. Rev. X}\ }\textbf {\bibinfo {volume} {2}},\ \bibinfo {pages}
  {011012} (\bibinfo {year} {2012})}\BibitemShut {NoStop}%
\end{thebibliography}

%

\onecolumngrid
\newpage
\appendix

\renewcommand{\thefigure}{\thesection\arabic{figure}}
\setcounter{figure}{0}

\centerline{\bf\large Supplemental Material}

\section{Non-linear response for slowly varying fields}
\label{Appendix A}

In this section we evaluate the quantum non-linear response formula, Eq. \ref{alphaw1w2} for potentials that vary slowly in space and time, and show that the result agrees with the perturbative solution of the Boltzmann equation, Eq. \ref{boltzmann}.  In the slowly varying limit, all of the transitions in (\ref{alphaw1w2}) will be intra-band, so a single band theory, characterized by dispersion $E({\bf k})$ will suffice.

Our starting point is to write Eq. \ref{alphaw1w2} in real space.
\begin{equation}
\alpha(\omega_1,\omega_2) = \frac{e^3}{\hbar^2}\int d^6{\bf r}_{1,2,3} Q_1({\bf r}_1) Q_2({\bf r}_2)Q_3({\bf r}_3) \chi(\left\{\omega_p,{\bf r}_p\right\})
\label{alphachi}
\end{equation}
with
\begin{equation}
\chi(\left\{{\bf r}_p,\omega_p\right\}) =
 \sum_{{\bf k}_{l,m,n}} 
 \frac{f_l-f_m}{\omega_1 -\omega_{lm}+i\eta}[
\frac{\rho_{lm}({\bf r}_1)\rho_{mn}({\bf r}_2)\rho_{nl}({\bf r}_3)}{\omega_1+\omega_2-\omega_{ln}+2i\eta}
 -\frac{\rho_{lm}({\bf r}_1)\rho_{mn}({\bf r}_3)\rho_{nl}({\bf r}_2)}{\omega_1+\omega_2-\omega_{nm}+2i\eta}] + (1\leftrightarrow 2).
\end{equation}
When $Q_p({\bf r}_p)$ is slowly varying, the matrix elements of the density operator will be dominated by intraband terms with small momenta, ${\bf q}_{lm} = {\bf k}_l-{\bf k}_m\sim 1/b$,
\begin{equation}
\rho_{lm}({\bf r}) = \langle {\bf k}_l|\rho({\bf r})|{\bf k}_m\rangle = e^{-i{\bf q}_{lm}\cdot{\bf r}} (1 + O({\bf q}_{lm})).
\end{equation}
Expanding to leading order in ${\bf q}_{lm}$ then gives
\begin{equation}
\chi(\left\{{\bf r}_p,\omega_p\right\}) =
 \sum_{{\bf k}_{l,n,m}} 
 \frac{f_l-f_m}
{\omega_1 - {\bf q}_{lm}\cdot{\bf v}_{lm}+i\eta}[
\frac{e^{i( {\bf q}_{lm}\cdot{\bf r}_{31} + {\bf q}_{nm}\cdot{\bf r}_{32})}}
{\omega_1+\omega_2-{\bf q}_{ln}\cdot{\bf v}_{ln}+2i\eta}
 -\frac{e^{i( {\bf q}_{lm}\cdot{\bf r}_{31} + {\bf q}_{ln}\cdot{\bf r}_{32})}}
{\omega_1+\omega_2-{\bf q}_{nm}\cdot{\bf v}_{nm}+2i\eta}] + (1\leftrightarrow 2)
\end{equation}
where since it is necessary to keep order $q^2$ terms in the denominator we write 
$\omega_{lm} = {\bf q}_{lm} \cdot {\bf v}_{lm} + O(q^3)$, with 
${\bf v}_{lm} = {\bf v}_{({\bf k}_l + {\bf k}_m)/2}$.   
If we define ${\bf k} = ({\bf k}_l + {\bf k}_m)/2$, 
then
${\bf v}_{lm} = {\bf v}_{\bf k} $, 
${\bf v}_{ln} = {\bf v}_{{\bf k} + {\bf q}_{nm}/2}$ and
${\bf v}_{nm} = {\bf v}_{{\bf k} - {\bf q}_{ln}/2}$.
We next define ${\bf q}_1 = {\bf q}_{lm}$ and ${\bf q}_2 = {\bf q}_{mn}$ (${\bf q}_{ln}$) in the first (second) term.  It follows that
\begin{align}
\chi(\left\{{\bf r}_p,\omega_p\right\}) =
  &\sum_{{\bf k},{\bf q}_1,{\bf q}_2} 
e^{i({\bf q}_1 \cdot {\bf r}_{31} + {\bf q}_2 \cdot {\bf r}_{32})}
 \frac{f({\bf k}+ {\bf q}_1/2)-f({\bf k}-{\bf q}_1/2) }
{\omega_1 - {\bf q}_1\cdot{\bf v}_{\bf k}+i\eta} \nonumber\\
&[
\frac{1}
{\omega_1+\omega_2-({\bf q}_1+{\bf q}_2)\cdot{\bf v}_{{\bf k} - {\bf q}_2/2}+2i\eta}
 -\frac{1}
{\omega_1+\omega_2-({\bf q}_1+{\bf q}_2)\cdot{\bf v}_{{\bf k} +{\bf q}_2/2}+2i\eta}] + (1\leftrightarrow 2)\\
 =
 &-\sum_{{\bf k},{\bf q}_1,{\bf q}_2} 
e^{i({\bf q}_1 \cdot {\bf r}_{31} + {\bf q}_2 \cdot {\bf r}_{32})}
 \frac{{\bf q}_1 \cdot\nabla_{\bf k}f({\bf k})  }
{\omega_1 - {\bf q}_1\cdot{\bf v}_{\bf k}+i\eta} 
{\bf q}_2\cdot\nabla_{\bf k}
\frac{1}
{\omega_1+\omega_2-({\bf q}_1+{\bf q}_2)\cdot{\bf v}_{{\bf k} }+2i\eta}
+ (1\leftrightarrow 2)\\
 =
 &\nabla^a_{{\bf r}_1} \nabla^b_{{\bf r}_2} \sum_{{\bf k},{\bf q}_1,{\bf q}_2} 
 e^{i({\bf q}_1 \cdot {\bf r}_{31} + {\bf q}_2 \cdot {\bf r}_{32})}
\frac{\nabla_{\bf k}^a f({\bf k}) }
{\omega_1 - {\bf q}_1\cdot{\bf v}_{\bf k}+i\eta} 
\nabla_{\bf k}^b
\frac{1}
{\omega_1+\omega_2-({\bf q}_1+{\bf q}_2)\cdot{\bf v}_{{\bf k} }+2i\eta}
+ (1\leftrightarrow 2)
\end{align}
Now we can do the sums over ${\bf q}_1$ and ${\bf q}_2$ using independent variables ${\bf q}_1$ and ${\bf q}_3 = -{\bf q}_1 - {\bf q}_2$.   Let us define
\begin{equation}
D({\bf r},\omega,{\bf v}) = i\sum_{\bf q} \frac{e^{i {\bf q}\cdot {\bf r}}}{\omega - {\bf q}\cdot {\bf v}+i\eta} 
=  \frac{e^{(i \omega-\eta) |{\bf r}|/|{\bf v}|}}{|{\bf v}|} \delta(\hat {\bf v} \times {\bf r}) \theta(\hat {\bf v}\cdot {\bf r})
\label{Deq}
\end{equation}
Then
\begin{equation}
\chi(\left\{{\bf r}_p,\omega_p\right\}) =
-\nabla^a_{{\bf r}_1} \nabla^b_{{\bf r}_2} \sum_{{\bf k}}  (\nabla^a_{\bf k} f({\bf k}))
D({\bf r}_{21},\omega_1,{\bf v}_{\bf k}) \nabla^b_{\bf k} D({\bf r}_{32},\omega_1+\omega_2,{\bf v}_{\bf k}) 
+ (1\leftrightarrow 2)   
\label{chifinal}
\end{equation}
Plugging (\ref{chifinal}) into (\ref{alphachi}) reproduces Eq. \ref{finalalpha} in the main text.   Note, that the same result would follow by integrating the Boltzmann equation (Eq. \ref{boltzmann}) to order $V_1 V_2$, with ${\bf E}_{p=1,2}({\bf r},t) = -\nabla Q_p({\bf r}) V_p e^{(\eta-i\omega_p)t}$.

\section{Non-linear response for a triple point contact}

\label{Appendix B}

\begin{figure}
\includegraphics[width=5in]{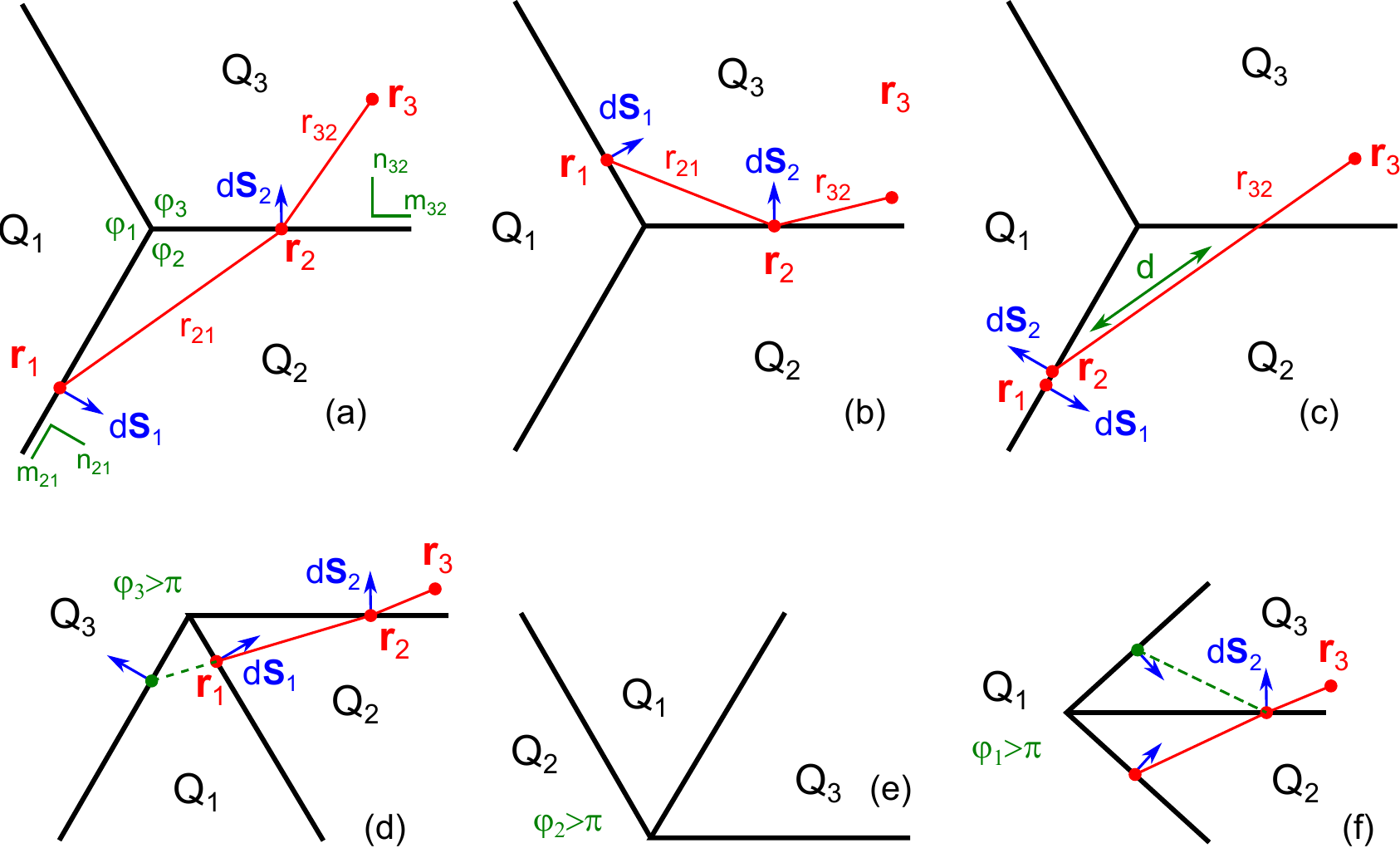}
\caption{A triple point contact with three contact regions that meet at a point and subtend arbitrary angles $\varphi_{1,2,3}$.   (a,b,c) show three contributions to $\alpha_1(\omega_1,\omega_2)$ for the case all angles $\varphi_p < \pi$.  (a) and (b) contribute to the intrinsic term, while (c) contributes to the extrinsic term.   (d,e,f) show contributions to the intrinsic term in the cases where one of the angles is greater than $\pi$.   For (d) and (e) the contribution is zero, while for (f) it is the same as (a) and (b).}
\label{figb1}
\end{figure}

In this appendix we evaluate the non-linear response in the frequency domain described by Eq. \ref{finalalpha} for a triple point contact, depicted in Fig. \ref{figb1} in which the three contact regions meet at a point and subtend arbitrary angles $\varphi_{p=1,2,3}$.   We will show that the non-linear response function consists of two terms with distinct frequency dependences.  There is an intrinsic term of the form, 
\begin{equation}
\alpha_i(\omega_1,\omega_2) = \chi_F\frac{e^3}{h^2}\frac{1}{(\eta-i\omega_1)(\eta-i\omega_2)}
\label{alphaint}
\end{equation}
which depends only on the Euler characteristic of the Fermi sea and is insensitive to $\varphi_p$, provided all three angles satisfy $\varphi_p<\pi$.   This generalizes the analysis of Eqs. \ref{boltzmann}-\ref{q3chiF} to the frequency domain and to a more general geometry.  At the end we will show how this result is modified when one of the contact angles is greater than $\pi$.
 In addition, there is an extrinsic term,
\begin{equation}
\alpha_e(\omega_1,\omega_2) = k \frac{e^3}{h^2}\frac{1}{(\eta-i(\omega_1+\omega_2))^2}
\label{alphaext}
\end{equation}
where $k$ is a dimensionless constant that depends on the details of the Fermi surface, as well as the angles $\varphi_p$.   An explicit formula for $k$ will be given below.

We assume the region boundaries, where $\nabla Q_p({\bf r})$ is non-zero, are confined a region of width $b$ about the straight rays shown in Fig. \ref{figb1}.   The slowly varying condition is satisfied provided $k_F b \gg 1$.  Provided $\omega b \ll 1$, the result will be independent of $b$.
Due to the gradients on $Q_p({\bf r}_p)$ for $p=1,2$ in Eq. \ref{finalalpha}, ${\bf r}_{p=1,2}$ are confined to the boundary of region $p$.   We thus evaluate 
\begin{equation}
\alpha(\omega_1,\omega_2) = \alpha_1(\omega_1,\omega_2) + \alpha_2(\omega_1,\omega_2)
\end{equation}
with
\begin{equation}
\alpha_1(\omega_1,\omega_2) = \frac{e^3}{\hbar^2} 
\sum_{\bf k}\int d{\bf S}^a_1 d{\bf S}^b_2 d^2{\bf r}_3 
(\nabla_{\bf k}^a f({\bf k})) D({\bf r}_{21},{\bf k},\omega_1) \nabla_{\bf k}^b D({\bf r}_{32},{\bf k},\omega_1+\omega_2)
\label{alphasupp}
\end{equation}
where ${\bf r}_{p=1,2}$ are integrated along the boundary of region $p$, with the perpendicular length element $d{\bf S}_p$ pointing away from region $p$.   
$\alpha_2$ is obtained by interchanging $1 \leftrightarrow 2$.
 It is convenient to write \ref{Deq} as
\begin{equation}
D({\bf r},{\bf k},\omega) = \int_0^\infty dt e^{i\omega t} \delta^2({\bf r} - {\bf v}_{\bf k} t).
\end{equation}
We then obtain
\begin{equation}
\alpha_1(\omega_1,\omega_2) = \frac{e^3}{\hbar^2}
\int_0^\infty dt_{21} dt_{32} e^{i \omega_1 t_{21} + i (\omega_1+\omega_2)t_{32}}
\int d{\bf S}^a_1 d{\bf S}^b_2 d^2{\bf r}_3  
\sum_{\bf k}
(\nabla_{\bf k}^a f({\bf k})) \delta^2({\bf r}_{21}-{\bf v}_{\bf k}t_{21}) \nabla_{\bf k}^b 
\delta^2({\bf r}_{32}-{\bf v}_{\bf k}t_{32}).
\label{alpha1}
\end{equation}
This shows that the integrals are dominated by straight lines parallel to ${\bf v}_{\bf k}$ that begin on ${\bf r}_{1(2)}$ on the boundary of region 1(2) and end on ${\bf r}_3$ in the interior of region 3, while visiting ${\bf r}_{2(1)}$ on the boundary of region 2(1) along the way.   
Due to the derivative $\nabla_{\bf k}^b$ in (\ref{alpha1}) the two segments of the line should be considered to have slightly different slopes.   This is important when ${\bf v}_{\bf k}$ is parallel to one of the boundaries.

There are three contributions to $\alpha_1$: $\alpha_{1A,1B,1C}$, shown in panels (a,b,c) of Fig. \ref{figb1}, that depend on which segments of the boundaries ${\bf r}_1$ and ${\bf r}_2$ are on.  (A fourth combination is not present because a straight line ${\bf r}_1 \rightarrow {\bf r}_2 \rightarrow {\bf r}_3$ can not be formed.)   We will see that the intrinsic term is $\alpha_i = \alpha_{1A} + \alpha_{1B} + \alpha_{2A} + \alpha_{2B}$ and the extrinsic term is $\alpha_e = \alpha_{1C} + \alpha_{2C}$.

We parameterize the boundary ray separating region $p$ and region $q$ by writing ${\bf r}_p = z_p \hat m_{qp}$, where $0<z_p<\infty$ and $\hat m_{pq}\equiv\hat m_{qp}$ is a unit vector pointing along the ray.   The perpendicular length element is given by $d{\bf S}_p = dz_p\hat n_{pq}$, where $\hat n_{qp}\equiv -\hat n_{pq}$ is a unit vector perpendicular to $\hat m_{pq}$ pointing away from region $p$.  

\subsection{Intrinsic Terms}

For $\alpha_{1A}(\omega_1,\omega_2)$ (Fig. \ref{figb1}a), the integral over ${\bf r}_3$ can be evaluated using
\begin{equation}
\int d^2{\bf r}_3 \delta^2({\bf r}_3 - z_{32} \hat m_{32} - {\bf v}_{\bf k} t_{32}) = \theta({\bf v}_{\bf k} \cdot \hat n_{32}),
\end{equation}
where we note that $t_{32}>0$.  The integral over ${\bf r}_{1,2}$ follows from
\begin{equation}
\int_0^\infty dz_1 dz_2 \delta^2( z_2 \hat m_{32} - z_1 \hat m_{21} - {\bf v}_{\bf k} t_{21})
= \frac{\theta({\bf v}_{\bf k}\cdot \hat n_{21}) \theta({\bf v}_{\bf k}\cdot \hat n_{32})}{|\hat m_{32}\times \hat m_{21}|}.
\label{r12integral}
\end{equation}
We then find
\begin{equation}
\alpha_{1A}(\omega_1,\omega_2) = \frac{e^3}{\hbar^2}
\int_0^\infty dt_{21} dt_{32} e^{i\omega_1 t_{21} + i(\omega_1+\omega_2)t_{32}}
\sum_{\bf k} (\hat n_{21}\cdot \nabla_{\bf k} f({\bf k}))
\frac{\theta({\bf v}_{\bf k}\cdot \hat n_{21})\theta({\bf v}_{\bf k}\cdot \hat n_{32})}{|\hat m_{32}\times \hat m_{21}|}
\hat n_{32}\cdot \nabla_{\bf k} \theta({\bf v}_{\bf k} \cdot \hat n_{32}) 
\end{equation}
Since $\nabla_{\bf k} \theta({\bf v}_{\bf k} \cdot \hat n_{32})$ fixes ${\bf v}_{\bf k} \parallel \hat m_{32}$, and $\nabla_{\bf k} f({\bf k}) \parallel {\bf v}_{\bf k}$ we can write
$\hat n_{21}\cdot \nabla_{\bf k} f({\bf k}) = (\hat n_{21} \cdot \hat m_{32}) \hat m_{32} \cdot\nabla_{\bf k} f({\bf k})$, along with $\theta({\bf v}_{\bf k}\cdot \hat n_{21}) = 
\theta({\bf v}_{\bf k} \cdot m_{32})$.  Since $\hat n_{21}=\hat z \times \hat m_{21}$, 
$\hat n_{21} \cdot \hat m_{32}/|\hat m_{32}\times \hat m_{21}|=1$, and we obtain,
\begin{equation}
\alpha_{1A}(\omega_1,\omega_2) =  
\frac{e^3/\hbar^2}{(\eta-i(\omega_1+\omega_2))(\eta-i\omega_1)}
\sum_{\bf k} (\hat m_{32} \cdot\nabla_{\bf k} f({\bf k}))
\theta(\hat m_{32}\cdot {\bf v}_{\bf k}) 
\theta(\hat n_{32}\cdot {\bf v}_{\bf k})
\hat n_{32}\cdot \nabla_{\bf k} 
\theta({\bf v}_{\bf k} \cdot \hat n_{32}) 
\end{equation}

The analysis of $\alpha_{1B}(\omega_1,\omega_2)$ is almost the same.   The only difference is that in (\ref{r12integral}) $\theta({\bf v}_{\bf k}\cdot \hat n_{32})$ is replaced by
$\theta(-{\bf v}_{\bf k}\cdot \hat n_{32})$.   For $\alpha_{1A}+\alpha_{1B}$ we can use
$\theta(x)+\theta(-x)=1$.  After adding a similar contributions to $\alpha_{2A}+\alpha_{2B}$ we obtain the intrinsic term in the non-linear response 
\begin{equation}
\alpha_i(\omega_1,\omega_2) =  
\frac{e^3/\hbar^2}{(\eta-i(\omega_1+\omega_2))(\eta-i\omega_1)}
\sum_{\bf k} (\hat m_{32} \cdot\nabla_{\bf k} f({\bf k}))
\theta(\hat m_{32}\cdot {\bf v}_{\bf k}) 
\hat n_{32}\cdot \nabla_{\bf k} 
\theta({\bf v}_{\bf k} \cdot \hat n_{32}) + (1\leftrightarrow 2)
\end{equation}
Following the analysis of Eq. \ref{q3fs}, the sum on ${\bf k}$ gives $\chi_F/(2\pi)^2$, and combining the two frequency dependent terms leads to (\ref{alphaint})

We next consider the contributions to $\alpha_i$ in the case where one of the angles $\varphi_p$ is greater than $\pi$.   The case $\alpha_3 >\pi$ is shown in Fig \ref{figb1}a.   In that case the contributions due to the two boundaries of region 1 (indicated by the red and green dots) have opposite sign due to the opposite projections of $d{\bf S}_1$ onto ${\bf v}_{\bf k}$, and in fact, they exactly cancel:  $\alpha_{1A}+\alpha_{1B} = 0$.    In the case $\alpha_2 > \pi$, shown in Fig. \ref{figb1}b, there are no contributions, since there are straight lines visiting regions $ 1\rightarrow 2 \rightarrow 3$, so again, $\alpha_{1A}+\alpha_{1B} = 0$.   On the other hand, when $\varphi_1>\pi$, shown in Fig. \ref{figb1}c, the contributions from the two boundaries of region 1 have the same sign, and the analysis is identical to the analysis when $\varphi_1<\pi$.

Combining these results, along with corresponding results for $\alpha_{2A}+\alpha_{2B}$, the general formula for the intrinsic term is,
\begin{equation}
\alpha_i(\omega_1,\omega_2) = \frac{\chi_F e^3/h^2}{2\eta-i(\omega_1+\omega_2)}\left[
\frac{\xi_2 \xi_3}{\eta-i\omega_1} + \frac{\xi_1 \xi_3}{\eta-i\omega_2}\right]
\label{alphawithxi}
\end{equation}
where 
\begin{equation}
\xi_p = \left\{\begin{array}{ll} 1 & \varphi_p <\pi \\ 0 & \varphi_p > \pi \end{array}\right.
\end{equation}
For the case where all angles are less than $\pi$ we recover Eq. 1.
Note that for the special case treated in Eq. 6-10, with $\varphi_1=\pi$, when the $V_2$ pulse follows the $V_1$ pulse, the excess charge $Q_3$ depends only on the first term in the brackets of (\ref{alphawithxi}), and is insensitive to $\varphi_1=\pi$.

\subsection{Extrinsic Terms}

We now consider the contribution $\alpha_{1C}$ from Fig. \ref{figb1}c.   This term did not show up in the time-domain pulse argument presented in the main text because there we assumed that $t_2>t_1$, so the ${\bf E}_1$ and ${\bf E}_2$ pulses were separated in time.   The frequency domain response, however, also includes a contribution from $t_1 \sim t_2$.   For the $\alpha_{1C}$ term, ${\bf r}_1 \sim {\bf r}_2$, so $t_1 \sim t_2$.   This leads to an extrinsic contribution that depends on the angles $\varphi_p$ as well as the shape of the Fermi surface, but has a distinct frequency dependence from the intrinsic term.

The integral over ${\bf r}_3$ gives,
\begin{equation}
\int d^2{\bf r}_3 \delta^2({\bf r}_3 - z_2 \hat m_{21} - {\bf v}_{\bf k} t_{32})
= \theta({\bf v}_{\bf k}\cdot\hat n_{32}) \theta(t_{32} - t_0(z_2))
\end{equation}
where the lower bound on $t_{32}$ is determined by the length $d$ of the segment in region 2,
\begin{equation}
t_0(z_2) = \frac{d}{|{\bf v}_{\bf k}|} = z_2 \frac{|\hat m_{21}\times \hat m_{32}|}{{\bf v}_{\bf k} \cdot \hat n_{32}}
\end{equation}
It is then useful to integrate the other $\delta$ function over $z_1$ and $t_{21}$:
\begin{equation}
\int dz_1 dt_{21} \delta^2( (z_2-z_1) \hat m_{21} - {\bf v}_{\bf k} t_{21}) =
 \frac{\theta({\bf v}_{\bf k}\cdot\hat n_{21})}{|\hat m_{21}\times {\bf v}_{\bf k}|}.
\end{equation}
This fixes $t_{21}=0$.   Expressing the remaining integral over $z_2$ as an integral over $t_0 = z_2 |\hat m_{21}\times \hat m_{32}|/{\bf v}_{\bf k} \cdot \hat n_{32}$ we find
\begin{equation}
\alpha_{1C}(\omega_1,\omega_2) = \frac{e^3}{\hbar^2}
\int_0^\infty dt_0 \int_{t_0}^\infty dt_{32} e^{i(\omega_1+\omega_2)t_{32}} 
\sum_{\bf k} (\hat n_{21}\cdot\nabla_{\bf k} f({\bf k}))
 \frac{\theta({\bf v}_{\bf k}\cdot\hat n_{21})}{|\hat m_{21}\times{\bf v}_{\bf k}|}
\hat n_{12}\cdot\nabla_{\bf k} \frac{{\bf v}_{\bf k} \cdot \hat n_{32}
\theta({\bf v}_{\bf k}\cdot\hat n_{32})}{|\hat m_{21}\times \hat m_{32}|}.
\end{equation}
Noting that $\nabla_{\bf k}f({\bf k}) = -{\bf v}_{\bf k} \delta(E_{\bf k})$, $\hat n_{12}=-\hat n_{21}$ and $\hat n_{21}\cdot{\bf v}_{\bf k}/||\hat m_{21}\times{\bf v}_{\bf k}|=1$ this becomes
\begin{equation}
\alpha_{1C}(\omega_1,\omega_2) = \frac{e^3}{\hbar^2}\frac{1}{(\eta-i(\omega_1+\omega_2))^2}
\sum_{\bf k} \delta(E_{\bf k}) 
\theta({\bf v}_{\bf k}\cdot\hat n_{21})\theta({\bf v}_{\bf k}\cdot\hat n_{32})
\frac{(\hat n_{21}\cdot\nabla_{\bf k})(\hat n_{32}\cdot\nabla_{\bf k})E_{\bf k}}
{|\hat n_{21}\times \hat n_{32}|}.
\end{equation}
Combining this with the corresponding term for $\alpha_{2C}$ leads to Eq. \ref{alphaext}, with the dimensionless constant $k$ given by
\begin{equation}
k = \int d^2{\bf k} \delta(E_{\bf k})\left[ 
\theta({\bf v}_{\bf k}\cdot\hat n_{21})\theta({\bf v}_{\bf k}\cdot\hat n_{32})
\frac{(\hat n_{21}\cdot\nabla_{\bf k})(\hat n_{32}\cdot\nabla_{\bf k})E_{\bf k}}
{|\hat n_{21}\times \hat n_{32}|} +
\theta({\bf v}_{\bf k}\cdot\hat n_{12})\theta({\bf v}_{\bf k}\cdot\hat n_{31})
\frac{(\hat n_{12}\cdot\nabla_{\bf k})(\hat n_{31}\cdot\nabla_{\bf k})E_{\bf k}}
{|\hat n_{12}\times \hat n_{31}|}
\right]
\end{equation}
$k$ involves an integral over the segment of the Fermi surface that has a velocity that points into region 3, and depends on the specific shape of the Fermi surface.

\end{document}